# Tailoring superconductivity in large-area single-layer NbSe$_2$ via self-assembled molecular adlayers

*Francesco Calavalle,[†] Paul Dreher,[#] Ananthu P. Surdendran,[§] Wen Wan,[#] Melanie Timpel,[∥] Roberto Verucchi,[∥] Celia Rogero,[‡,#] Thilo Bauch,[§] Floriana Lombardi,[§] Fèlix Casanova,[†,⊥] Marco Vittorio Nardi,[∥] Miguel M. Ugeda,[‡,⊥,#] Luis E. Hueso,*[†,⊥] and Marco Gobbi*[†,‡,⊥]*

[†]CIC nanoGUNE BRTA, Donostia-San Sebastian, Basque Country, 20018, Spain

[‡]Materials Physics Center CSIC-UPV/EHU, 20018 Donostia-San Sebastian, Spain

[§]Quantum Device Physics Laboratory, Department of Microtechnology and Nanoscience, Chalmers University of Technology, Göteborg, SE-41296, Sweden

[∥]Institute of Materials for Electronics and Magnetism, IMEM-CNR, Trento unit c/o Fondazione Bruno Kessler, Via alla Cascata 56/C, Povo, Trento IT-38123, Italy

[⊥]IKERBASQUE, Basque Foundation for Science, Bilbao, Basque Country, 48013, Spain

[#]Donostia International Physics Center DIPC, Donostia-San Sebastian, Basque Country, 20018, Spain

KEYWORDS

tunable superconductivity, NbSe2, transition metal dichalcogenide, large-area functionalization, self-assembly, monolayer TMD.

ABSTRACT

2D transition metal dichalcogenides (TMDs) represent an ideal testbench for the search of materials by design, since their optoelectronic properties can be manipulated through surface engineering and molecular functionalization. However, the impact of molecules on intrinsic physical properties of TMDs, such as superconductivity, remains largely unexplored. In this work, the critical temperature ($T_C$) of large-area $NbSe_2$ monolayers is manipulated employing ultra-thin molecular adlayers. Spectroscopic evidences indicate that aligned molecular dipoles within the self-assembled layers act as a fixed gate terminal, collectively generating a macroscopic electrostatic field on $NbSe_2$. This results in a ~55% increase and a 70% decrease in $T_C$ depending on the electric field polarity, which is controlled via molecular selection. The reported functionalization, which improves the air stability of $NbSe_2$, is efficient, practical, up-scalable and suited to functionalize large-area TMDs. Our results indicate the potential of hybrid 2D materials as a novel platform for tunable superconductivity.

TEXT

Transition metal dichalcogenides (TMDs) are layered compounds which can be thinned down to the single-layer limit.[1,2] While mechanical exfoliation generates atomically thin TMD flakes possessing an area of a few square microns, chemical and physical methods[3] provide high-quality monolayers on large-area substrates, which are suitable for actual technological applications. [3–8] Similar to other two-dimensional materials, TMD monolayers are characterized by extreme surface sensitivity,[9] making it possible to finely tune their optoelectronic properties through electrostatic gating or surface treatments.[10–12] Molecular functionalization is one of the most promising methods to engineer TMDs,[13–18] since an accurate choice of convenient functional groups makes it possible to provide programmable doping levels and unique responsivity to light and magnetic fields.[12,19–26] While most studies on molecular functionalization of TMDs demonstrate the engineering of the optoelectronic properties of micron-sized mechanically exfoliated flakes,[12,13,22–26,14–21] only few works focus on technologically relevant large-area TMDs,[27,28] leaving an open question about the up-scalability of chemical approaches. Moreover, the effect of organic adsorbates on other intrinsic properties of TMDs, like superconductivity, has been notably less explored.[13]

Despite its limited stability in air,[29] superconducting $NbSe_2$ has been intensely studied in the last decade because it exhibits intriguing electronic correlated phases.[6,8,29–32] In particular, $NbSe_2$ exhibits transition into a superconducting state below a critical temperature $T_C$,[33] which is lower in monolayers ($T_C \approx 1$ K) than in bulk crystals ($T_C \approx 7$ K).[6,30] The low-temperature superconducting state is gate tunable[29,34,35] and can be modified by molecular functionalization.[36–38] For instance, paramagnetic[37,38] and chiral molecules[39] were found to locally alter the surface superconductivity of bulk $NbSe_2$, giving rise to bound states within the superconducting gap, and signatures of ferromagnetism were detected in liquid phase exfoliated $NbSe_2$ interfaced with a polar reductive molecule.[36] However, a deterministic manipulation of the superconductivity of TMD monolayers by functionalization with on-purpose molecular adlayers has not yet been reported.

In this work we manipulate the critical temperature of large-area single-layer NbSe$_2$ in a deterministic way employing ultra-thin self-assembled adlayers. Functionalization with a fluorinated or an amine-containing molecule results in a 55% increase and a 70% decrease in the T$_C$ of NbSe$_2$ monolayers, respectively. We use ultraviolet photoemission spectroscopy (UPS) data to demonstrate that the recorded changes in T$_C$ are related to electric fields generated by the molecular adlayers, which act as an effective fixed gate terminal. Importantly, the polarity of the field-effect is determined by an accurate choice of appropriate functional groups. The presence of the ultrathin adlayer improves the air stability of NbSe$_2$, and the induced T$_C$ modification is only minimally affected by storing the sample in air for 60 hours. The relative variation in T$_C$ introduced by our chemical functionalization is larger than that reported for field-effect devices based on multilayer NbSe$_2$,[29,34,35] highlighting the high efficiency of molecular gating and the importance of employing TMDs at the monolayer limit for the modulation of their intrinsic properties.

RESULTS AND DISCUSSION

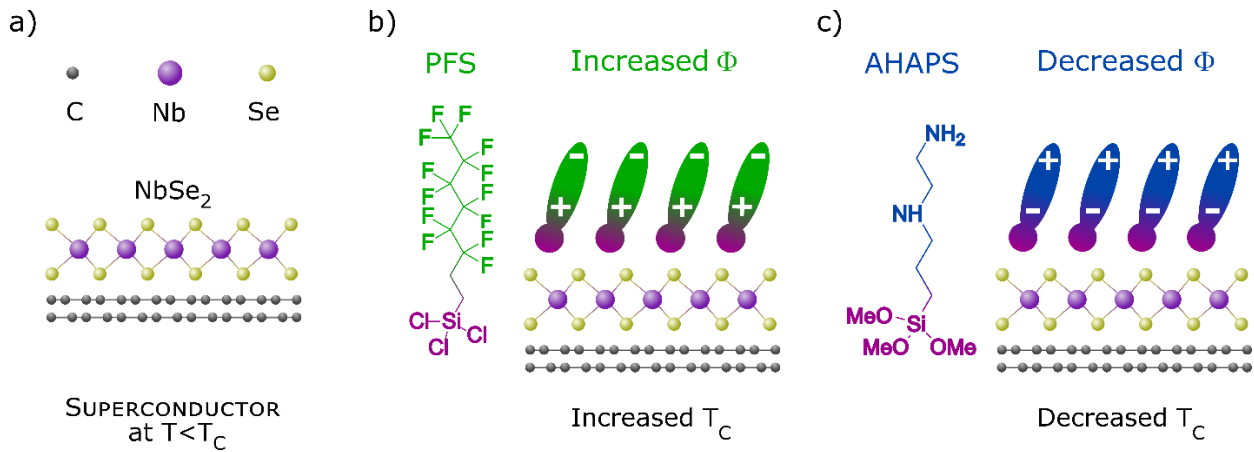

**Figure 1.** Schematic representation of our approach to manipulate the critical temperature Tc of NbSe$_2$. (a) large-area NbSe$_2$ monolayer grown on epitaxial bilayer graphene (NbSe$_2$/BLG/SiC); molecular functionalization of the NbSe$_2$ monolayer with (b) trichloro(1H,1H,2H,2H-

perfluorooctyl)silane (PFS) and (c) N-[3-(trimethoxysilyl)propyl]ethylenediamine (AHAPS). The direction of the PFS (AHAPS) molecule's permanent dipole moment leads to an increase (decrease) of the work function ($\Phi$), which in turn causes an increase (decrease) of $T_C$, due to hole (electron) accumulation.

A schematic representation of our approach is shown in **Figure 1**. We employed single-layer $NbSe_2$ grown by molecular beam epitaxy over mm-sized areas. To characterize the effect of molecules on the superconducting transition, we measured the $NbSe_2$ $T_C$ before and after functionalizing its surface with silane-containing molecules (**Figure 1**). The functionalization of the exposed surface of 2D Materials with molecules containing silane groups holds a great potential, since their self-assembly generates ultrathin and ordered films firmly attached to the 2D surface.[21–23,40,41] However, this approach has only been explored in a very limited number of works.[21–23,40,41]

We selected two linear molecules containing a silane group as anchoring group and a polar functional group, which are Trichloro(1H,1H,2H,2H-perfluorooctyl)silane (PFS) and N-[3-(trimethoxysilyl)propyl]ethylenediamine (AHAPS, Figure 1b, c). While PFS incorporates a fluorinated chain as polar functional group, AHAPS contains two amino groups.

Single-layer $NbSe_2$ was grown on epitaxial bilayer graphene (BLG) on 6H-SiC(0001), denoted as $SiC/BLG/NbSe_2$, for the electrical measurements and on the chemically analogous surface of highly oriented pyrolytic graphite (HOPG), denoted as $NbSe_2/HOPG$, for the spectroscopic measurements. The quality of our $NbSe_2$ films was characterized by Atomic force microscopy (AFM) and X-ray photoemission spectroscopy (XPS). **Figure 2**a shows the typical morphology of a $NbSe_2/BLG/SiC$ sample, as imaged by AFM. In this case, the $NbSe_2$ monolayer covers almost completely the terraces of epitaxial graphene on SiC. We note that, even if the coverage is not complete, long-ranged charge carrier percolation is achieved in these NbSe$_2$ layers, as proved by the fact that we measure a superconducting state with zero resistance (see below).

The samples grown on HOPG (Figure 2b) present a similar morphology, i.e., a nearly full monolayer although decorated with scattered islands of bilayer. In both cases, the thickness of NbSe$_2$ layer is 0.6 ± 0.1 nm (see insets), in agreement with the reported thickness of TMD monolayers.[33]

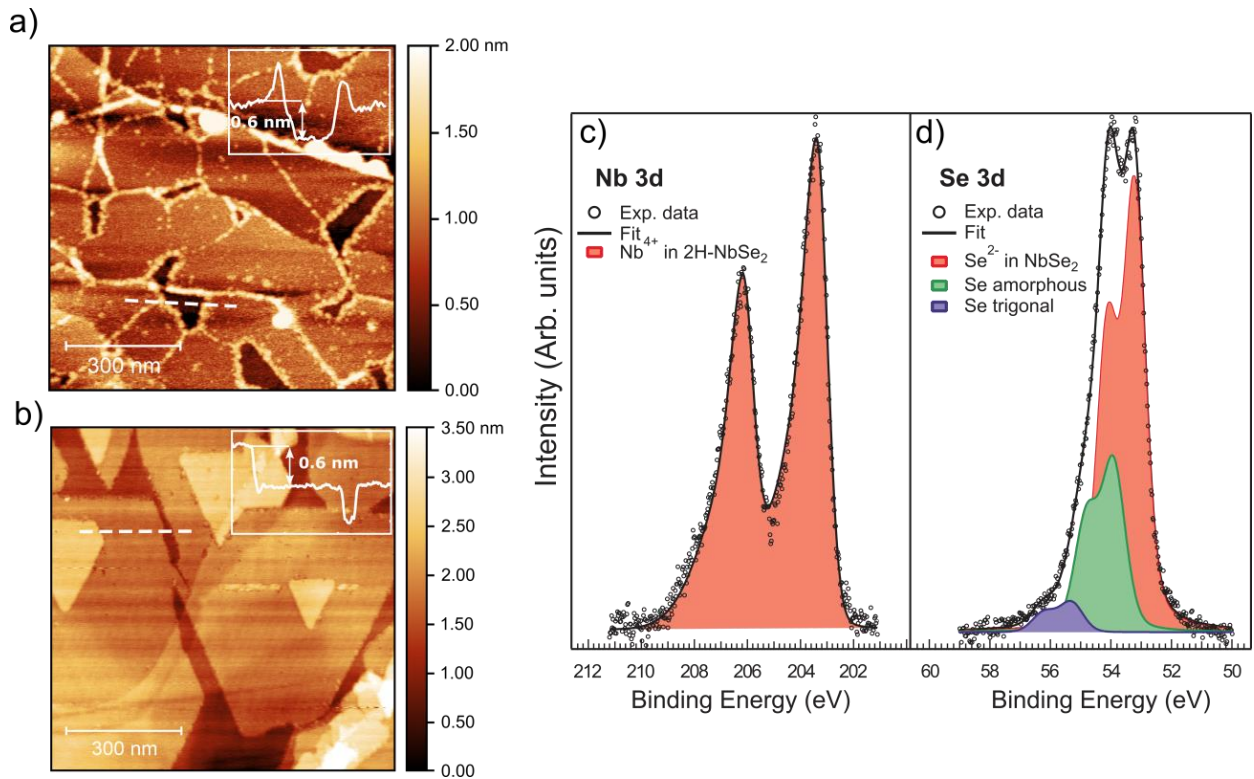

**Figure 2.** Morphological and spectroscopic characterization of NbSe$_2$ thin films grown by molecular beam epitaxy. a,b) AFM images of NbSe$_2$ grown on BLG/SiC and on HOPG, respectively. The insets show height profiles along the white lines in a) and b). c) Fitted Nb 3d and d) Se 3d core level XPS spectra of NbSe$_2$/BLG/SiC. The Se 3d core level spectrum in Figure 2d

can be deconvolved into three components, with the main component corresponding to $Se^{2-}$ in NbSe$_2$ (Se 3d$_{5/2}$ peak at 53.25 eV, red component). The two minor components can be ascribed to residual amorphous Se (green component at 53.95 eV) and trigonal Se (blue component at 55.35 eV).

Chemical analysis of the NbSe$_2$/BLG/SiC substrate was conducted by XPS. The fitted Nb 3d and Se 3d core level are displayed in Figure 2c and d, respectively. The spectral features of the Nb 3d peak indicate that Nb has a 4+ oxidation state, as expected for the Nb coordinated with 6 Se atoms in the NbSe$_2$ 2H hexagonal phase. In particular, the Nb 3d$_{5/2}$ and 3d$_{3/2}$ peaks stemming from Nb$^{4+}$ in the 2H phase of NbSe$_2$ (red components in Figure 2c) are located at BE = 203.40 and 206.15 eV, respectively. The Se 3d core level spectrum in Figure 2d can be deconvolved into three components, with the main component stemming from $Se^{2-}$ in NbSe$_2$.[42,43] The fitting procedure resulted in an atomic ratio Se/Nb of 2.1, in good agreement with the expected ratio of 2.0, corroborating the high quality of the NbSe$_2$ monolayers. The other two minor components can be ascribed to residual amorphous Se and trigonal Se, which form as clusters on the surface that are imaged as bright protrusions in large area AFM maps (see Figure S1).

In order to functionalize the NbSe$_2$ monolayers with self-assembled molecular adlayers, we simply exposed the NbSe$_2$ surface to vapors of PFS and AHAPS (see Methods in SI and Figure S2). A similar method was employed to functionalize the chemically similar van der Waals surface of other micron-sized 2D Materials with PFS, AHAPS and other silane-based molecules.[21–23,40,41]

In **Figure 3**, we present AFM images of NbSe$_2$/HOPG after functionalization with PFS (Figure 3a) and AHAPS (Figure 3b). The functionalized substrates display a markedly different morphology. NbSe$_2$ islands characterized by 0.6-nm-thick step edges can be discerned also after functionalization with AHAPS and PFS, but in both cases they are covered by thin and smooth molecular films characterized by a roughness of 0.3 ± 0.1 nm (Figures 3a, b). A similar morphology was also recorded in AFM images measured in different regions of the samples and separated by a few millimeters, indicating that the molecular adlayers extend homogenously over the entire substrate area. Vapor phase deposition is therefore ideal to functionalize large-area TMDs with homogeneous molecular adlayers.

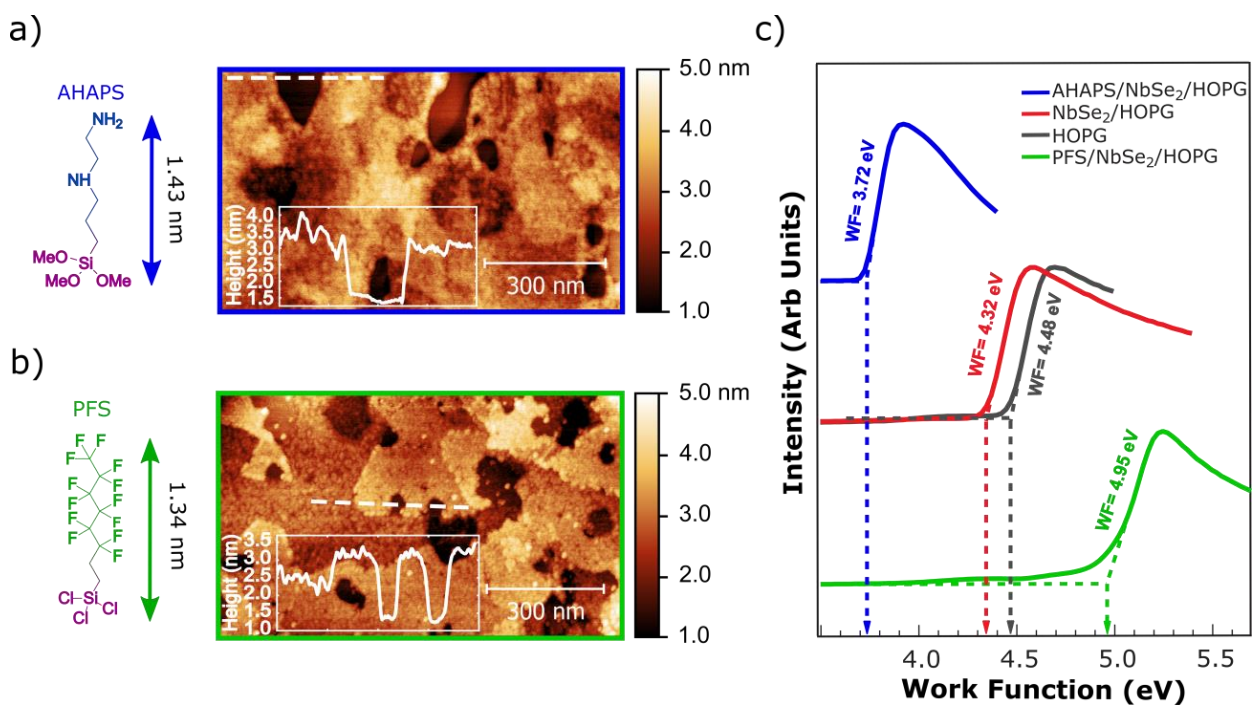

**Figure 3.** Morphological and spectroscopic characterization of AHAPS and PFS self-assembled adlayers on NbSe$_2$ monolayers on HOPG. a-c) AFM images of NbSe$_2$ treated with AHAPS a) and PFS b) adlayers. The insets show the thickness of the molecular adlayers, as extracted from AFM profiles, which can be compared with the length of the molecules. c) Work function of bare HOPG (black line), untreated NbSe$_2$ monolayers (red line), AHAPS-treated NbSe$_2$ (blue line) and PFS-treated NbSe$_2$ (green line), as extracted by the secondary electron cutoff (SECO) spectra.

The growth parameters of the molecular adlayers shown in Figures 3a and b were intentionally optimized to achieve a homogeneous but non-complete substrate coverage (Figure S2). In this way, AFM profiles measured across the uncovered areas were used to obtain information on the molecular layer thickness. Instead, the spectroscopic and electrical characterization (see below) was performed on fully covered samples to maximize the effect of the molecular adlayers on NbSe$_2$ (Figure S3).

We estimate a thickness of 1.8 ± 0.5 nm for the AHAPS layer and of 1.5 ± 0.3 nm for the PFS layer (inset in Figure 3a, b); both values are in good agreement with the reported length of the two linear molecules.[44] This finding suggests that AHAPS and PFS self-assemble generating ultrathin adlayers in which individual molecules are preferentially aligned in a direction roughly perpendicular to the NbSe$_2$ surface, in agreement with the reported formation of ordered assembly of silane molecules on WSe$_2$ and graphene.[21,40] Further evidences confirming the reported formation of the assembly come from the spectroscopic characterization, as discussed in Figure S4.

Importantly, the molecular ordering in the self-assembled adlayer implies that the permanent dipoles of AHAPS and PFS are aligned and possess a component in the direction orthogonal to the surface, which introduces a field effect on NbSe$_2$.[21,23,40] The effect of the molecular electric fields on NbSe$_2$ was characterized by means of UPS.[45] Initially, we characterized how the presence of monolayer NbSe$_2$ affects the work function ($\Phi$) of the HOPG substrate. As can be seen in Figure 3c, the secondary electron cut-off (SECO) position of the bare HOPG surface yields a typical $\Phi$ of 4.48 eV, whereas for HOPG/NbSe$_2$, $\Phi$ is slightly reduced to 4.32 eV.

Afterwards, we investigated how the $\Phi$ of single-layer NbSe$_2$ is modified after functionalization with PFS and AHAPS. Remarkably, the functionalization with PFS increases $\Phi$ by +0.63 eV, whereas AHAPS decreases it by -0.60 eV. Such a change of Φ originates from the electric field generated by the superposition of the permanent molecular dipoles in AHAPS and PFS, which are aligned in the direction perpendicular to the surface due to the preferential out-of-plane molecular orientation framed by self-assembly. The negligible changes in the position of the Nb and Se core level after functionalization (Table S1) excludes significant charge transfer from the molecular adlayer to NbSe$_2$.

The recorded Φ shifts provide further evidence of the presence of long-range structural order within the self-assembled adlayer. The fact that Φ increases or decreases after functionalization with PFS or AHAPS indicates that the polarity of the field effect is opposite in the two cases, implying that the molecular dipoles possess an opposite orientation in the two self-assembled adlayers. This finding further confirms the scenario in which the silanol groups in AHAPS and PFS lie close to the NbSe$_2$ surface.[21,40]

A shift in the Φ of 2D materials corresponds to a change in their charge carrier density.[46] In our case, the increased Φ in PFS-functionalized NbSe$_2$ corresponds to an increase in the hole density, while the electric field effect generated by the AHAPS adlayer introduces electron accumulation in NbSe$_2$, or a decrease in the hole density. From the Φ shift values we can provide a rough estimation of the density of charge carriers $\Delta n$ induced by molecular gating using a simple parallel-plate capacitor model in which the molecular adlayer acts as dielectric, using the following formula:

$$\Delta n = \varepsilon_0 \varepsilon_{eff} \frac{\Delta \Phi}{et}$$

Where $\varepsilon_0$ is the vacuum permittivity, $\varepsilon_{eff}$ is the effective dielectric constant of the molecular adlayer, $\Delta\Phi$ is the work function shift, t is the thickness of the molecular adlayer and e the electron charge. Employing the values of $\Delta\Phi$ and t measured through UPS and AFM, and $\varepsilon_{eff} = 3 - 4$ as a typical value for the effective dielectric constant of a molecular adlayer,[47,48] we can estimate that the PFS and AHAPS adlayers introduce a hole and electron accumulation in the range of $\Delta p \sim \Delta n \sim 5 \times 10^{12} - 1 \times 10^{13}$ cm$^{-2}$. Remarkably, such induced charge carrier density is close to the typical maximum value that can be achieved through electrostatic gating employing SiO$_2$ as a dielectric, indicating that the field effect generated by our ultrathin molecular film or gating across significantly thicker SiO$_2$ are comparable, thereby demonstrating the high efficiency of our molecular gating.

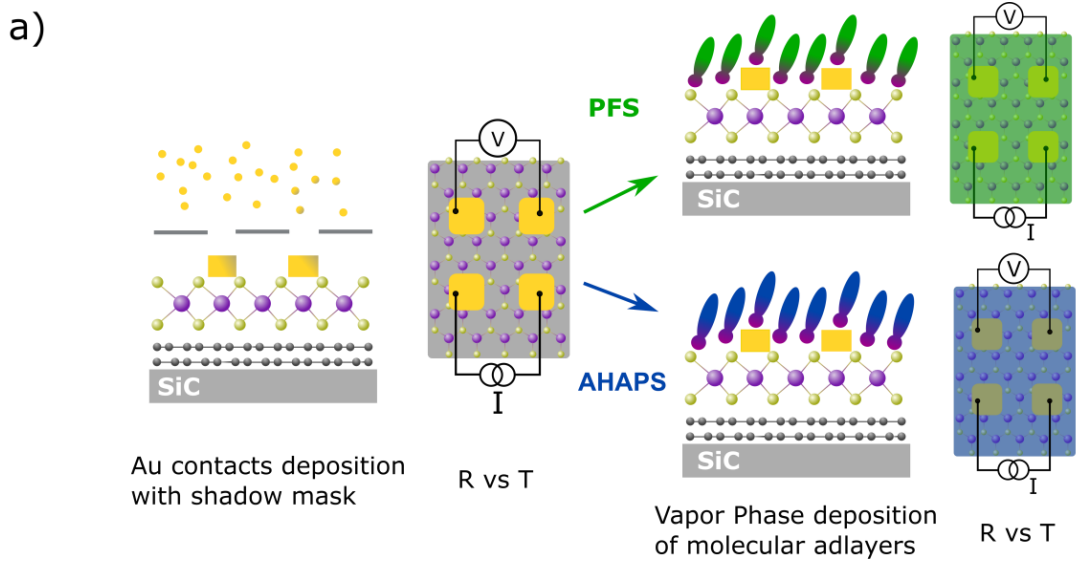

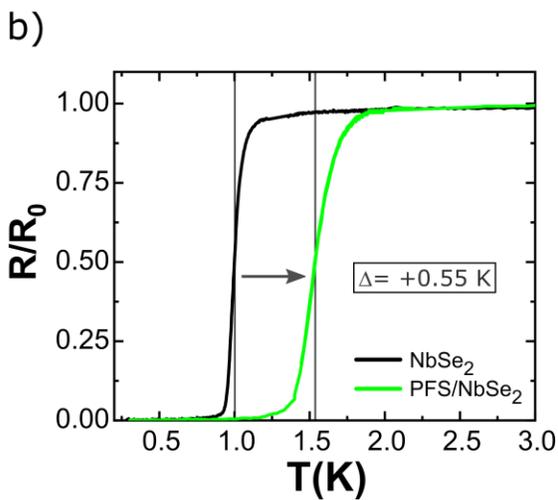

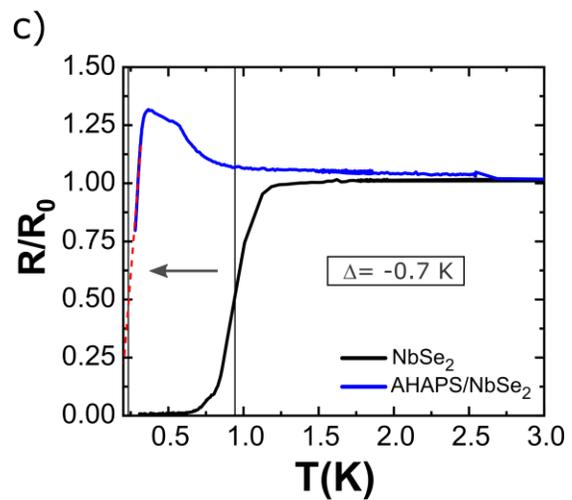

**Figure 4.** a) Scheme showing the experimental procedure employed to perform the electrical characterization. We first deposited four macroscopic gold electrodes by evaporation through shadow masks, and then we grew the molecular adlayer. The four-probe resistance was measured before and after the molecular functionalization. Normalized R vs T measurements of NbSe$_2$ samples before and after the deposition of b) PFS and c) AHAPS thin layers. In b) the T$_C$ increase from 1.00 K to 1.55 K while in c) the T$_C$ decreased from 0.90 K to a value lower than the limit of the measurement (0.29 K).

Previous works on NbSe$_2$ field-effect devices showed that a change in the charge carrier density of NbSe$_2$ translates in a variation of the electron-phonon interaction, crucial for the formation of the Cooper pairs and the superconductive state.[35] As a result, the T$_C$ is tunable by electrostatic gating. To explore the variation in T$_C$ introduced by our molecular gates, we performed transport measurements on large-area NbSe$_2$ monolayers on epitaxial graphene down to 0.29 K. **Figures 4**b and c show the temperature dependence of the resistance measured in two different samples before and after functionalization with PFS and AHAPS, respectively. The resistances are normalized to the value of the resistance state above the critical temperature (R$_0$), and we defined T$_C$ as the value of temperature corresponding to R$_0$/2. Figures 4 a and b show that the pristine NbSe$_2$ adlayers displayed the same T$_C$ before functionalization (T$_{C0}$ = 1.0 ± 0.1 K). The broadened width of the superconducting transition is a manifestation of the two-dimensional character of the superconductor, which is described by a Berezinskii–Kosterlitz Thouless-type (BKT) transition.[6] After functionalization with the p-type PFS self-assembled adlayer, T$_C$ increased to 1.55 K, corresponding to a ΔT$_C$ = + 0.55 K. In the case of functionalization with the n-type AHAPS adlayer, the T$_C$ decreased to 0.2 K (ΔT$_C$ = - 0.7 K). Note that in this case, the T$_C$ suppression allows us to observe an upturn in the resistance, which is normally not observed due to the onset of superconductivity. While a detailed explanation on the physical origin of such phenomenon goes beyond the scope of this work, we highlight that molecular functionalization provides access to conductance regimes that cannot be otherwise reached. The reproducibility of these results was confirmed by measurements performed on other two samples (see Figure S5). These results are in agreement with previous studies on NbSe$_2$ field-effect devices, where p-type (n-type) doping was reported to increase (suppress) T$_C$,[35] reinforcing the similarity between the molecular adlayer and a fixed gate terminal.

To verify whether the recorded T$_C$ shifts were due to the electric fields generated by the molecular adlayers, we exposed another sample to vapors of trichloro(octadecyl)silane (OTS), a molecule widely used to generate hydrophobic self-assembled monolayers. Since OTS does not feature a strong permanent dipole, it is not expected to induce strong electric fields on NbSe$_2$. The experiment revealed that the change in the T$_C$ of NbSe$_2$ before and after OTS functionalization is significantly lower than what measured for PFS and AHAPS (ΔT$_C$~0.1 K, see Figure S6). These data indicate that ΔT$_C$ scales with the sign and strength of the molecular permanent dipoles, not only supporting the molecule-induced electrostatic effect as the main cause of variation in T$_C$, but also providing a rational route to molecular design. In particular, by employing silane as anchoring groups and polar functional groups as source of electric fields, it is possible to modulate the T$_C$ of NbSe$_2$ in a deterministic way, thanks to the ordered nanoscale arrangement of silane-based molecules on TMDs. Moreover, we tested the air stability of an uncapped and a PFS-treated NbSe2 by measuring the temperature dependence of their four-point resistance before and after exposing them to air for 60 hours. For the uncapped sample, we measured a 50% resistance increase, indicative of a significant degradation of NbSe$_2$ in air. On the contrary, the resistance of the PFS-treated NbSe$_2$ increased only for a factor 3 %, demonstrating that the molecular adlayer acts as a capping layer and improves the air stability of NbSe$_2$ (Figure S7). Additionally, the effect of PFS on T$_C$ is only minimally modified by storing the sample in air for 60 hours (Figure S7). These results fully exploit the potential of silane- functionalization of TMD, demonstrating that they are ideal molecules not only to improve the performances of electronic devices, but also to engineer the intrinsic properties of atomically thin TMDs extending over large areas.

Finally, to evaluate the efficiency of the molecular adlayer in the manipulation of the T$_C$ of NbSe$_2$, we compare the shift measured in this work with that previously obtained in field effect devices.[35] Employing strontium titanate/hBN as a gate dielectric,[29] a ΔT$_C$ of approximately 50 mK was reported in bilayer NbSe$_2$, which amounts to approximately one tenth of the ΔT$_C$ measured in our work. Additionally, employing the p-type PFS adlayer we recorded an increase in T$_C$, ΔT$_C$ = + 0.55 K, which is close to the ΔT$_C$ = + 0.7 K measured using ionic liquid gating, which accumulates approximately a ten-times-larger charge carrier density (see also Table S2).[35]

The large T$_C$ modulation reported in this study can be partly explained by the fact that, unlike previous studies focusing on NbSe$_2$ multi-layers (≥ 2 layers), we employ monolayers which minimize screening effects and are thus more sensitive to any changes in their environment. In addition, our results indicate that while the molecular adlayers introduce an electrostatic effect on NbSe$_2$, they do not merely act as a gate dielectric, as they introduce other physico-chemical modification at the NbSe$_2$ surface which enhance the field effect. For instance, the modified chemical environment of defect states which accompanies the silane functionalization affects the T$_C$ shift, and likely contributes to improve the efficiency of molecular functionalization. Our results show that the electrostatic effect plays a primal role in the determination of the recorded shifts, but we anticipate that mastering the other simultaneous phenomena will permit a higher control on the intrinsic physical properties of TMDs.

CONCLUSION

In conclusion, we were able to tune the superconductive transition in monolayer NbSe$_2$ employing ultrathin self-assembled molecular adlayers. In particular, the T$_C$ ~ 1 K characteristic of pristine NbSe$_2$ monolayers could be lowered to 0.2 K and increased to 1.55 K by functionalization with self-assembled adlayers composed of two silane-based molecules with different dipolar substitutions. The T$_C$ variation is explained as a consequence of opposite electrostatic field effects generated by the aligned molecular dipoles in the ultrathin molecular adlayers, which act as a fixed gate terminal. Remarkably, the overall relative variation in T$_C$, which is above 120%, is larger than what was achieved in field-effect devices based on multilayer NbSe$_2$, highlighting the high efficiency of molecular functionalization and the importance of using high quality TMD monolayers for the engineering of their intrinsic properties. Our chemical functionalization, which does not require any expensive vacuum evaporator nor any high temperature process, is practical, scalable and perfectly suited to tune the physical properties of technologically relevant TMD monolayers extending over large areas. We also highlight that molecular functionalization is fully compatible with further gating in field-effect transistors. In this regard, our molecular functionalization provides a new controllable starting environment for NbSe$_2$, which can be programmed by molecular design and eventually further finely manipulated with a gate terminal. Therefore, by combining electrostatic gating and our molecular functionalization, one could in principle engineer precisely the material properties, and investigate doping regions that cannot be explored using conventional solid-state dielectrics.

We predict that our approach could be of great interest to manipulate deterministically other intrinsic physical properties of 2D materials which can be tuned by electrostatic gating, including their magnetism[49] and topology.[50]

ASSOCIATED CONTENT

**Supporting Information**

Figures S1−8, Table S1-2, and references. Morphological characterization of $NbSe_2$ monolayers grown on HOPG and SiC. Mechanism of growth of PFS and AHAPS on $NbSe_2$ surface investigated via XPS. AFM characterization of the self-assembled adlayers growth on different substrates and with different coverages of the samples. XPS of the core levels of $NbSe_2$ before and after the functionalization. Electrical measurements of more $NbSe_2$ samples before and after the functionalization and after exposition to atmosphere. Electrical measurements at low T of $NbSe_2$ functionalized with OTS. A methods section was included also in Note S1 of SI to give additional details on samples fabrication and characterization.

AUTHOR INFORMATION

**Corresponding Authors**

*E-mail: l.hueso@nanogune.eu

*E-mail: m.gobbi@nanogune.eu

**Author Contributions**

F.C, M.G. and L.E.H. conceived the experiment. P.D. and W.W. grew the NbSe$_2$ films, under the supervision of M.M.U. F.C. carried out the AFM study, optimized the fabrication of the devices and, with the help of A.S. and T.B., performed the electrical measurements. M.T., M.V.N. and R.V. performed the XPS and UPS experiments. F.C. and M.G. co-wrote the paper. All authors discussed the results and contributed to the interpretation of data as well as to editing the manuscript.

**Notes**

The authors declare no competing financial interest.

ACKNOWLEDGMENT

This work was supported by the European Union H2020 under the Marie Sklodowska-Curie Actions (766025-QuESTech and 748971-SUPER2D) and by the ERC Starting grant LINKSPM (Grant 758558); by "la Caixa" Foundation (ID 100010434), under the agreement LCF/BQ/PI19/11690017 and by the Spanish MICINN under the Maria de Maeztu Units of Excellence Programme (MDM-2016-0618), Project No. MAT2015-65159-R, MAT2017-88377-C2-1-R, RTI2018-094861-B-100 and PID2019-108153GA-I00.